\documentclass[reprint,amsmath,amssymb,aps,prb,amssymb,superscriptaddress,notitlepage, longbibliography]{revtex4-1}
\usepackage[utf8]{inputenc}
\usepackage{color}
\usepackage{amsmath, braket}
\usepackage{mathtools}
\usepackage{bm}
\usepackage{amsfonts}
\usepackage{dcolumn}
\usepackage{natbib}
\usepackage{bbold}
\usepackage{soul}
\usepackage[dvipsnames]{xcolor}
\usepackage{amssymb}
\usepackage[breaklinks=true,colorlinks,citecolor=blue,linkcolor=blue,urlcolor=blue]{hyperref}
\usepackage{graphicx}
\usepackage{appendix}

\usepackage{mathtools}
\begin{document}
\title{Layered Topological Antiferromagnetic Metal at Room Temperature - YbMn$_{2}$Ge$_{2}$} 
%: YbMn$_{2}$Ge$_{2}$}
%\title{Antiferromagnetic metal YbMn$_{2}$Ge$_{2}$ \& it's topology}
\thanks{NJ and AC contributed equally to this project}
\author{Nirmalya Jana} 
%\thanks{NJ and AC contributed equally to this project}
\email{nirmalyaj20@iitk.ac.in}
\affiliation{Department of Physics, Indian Institute of Technology Kanpur, Kanpur-208016, India}
\author{Atasi Chakraborty} 
%\thanks{NJ and AC contributed equally to this project}
%\thanks{NJ and AC contributed equally to this project}
\email{atasi.chakraborty@uni-mainz.de}
\affiliation{Institut f\"{u}r Physik, Johannes Gutenberg Universit\"{a}t Mainz, D-55099 Mainz, Germany}
\author{Anamitra Mukherjee}
\email{anamitra@niser.ac.in}
\affiliation{School of Physical Sciences, National Institute of Science Education and Research, a CI of Homi Bhabha National Institute, Jatni 752050, India} 
\author{Amit Agarwal} 
\email{amitag@iitk.ac.in}
\affiliation{Department of Physics, Indian Institute of Technology Kanpur, Kanpur-208016, India}
%\date{\today}

\begin{abstract}
Metallic antiferromagnets are essential for efficient spintronic applications due to their fast switching and high mobility, yet room-temperature metallic antiferromagnets are rare. Here, we investigate YbMn$_2$Ge$_2$, a room temperature antiferromagnet, and establish it as an exfoliable layered metal with altermagnetic surface states. Using multi-orbital Hubbard model calculations, we reveal that its robust metallic AFM ordering is stabilised by electronic correlations and a partially nested Fermi surface. Furthermore, we show that YbMn$_2$Ge$_2$ hosts symmetry-protected topological Dirac crossings, connecting unique even-order spin-polarised surface states with parabolic and inverted Mexican-hat-like dispersion. Our findings position YbMn$_2$Ge$_2$ as a promising platform for exploring the interplay of correlation, topology, and surface altermagnetism of layered antiferromagnets.
\end{abstract}

\maketitle 
%\textcolor{red}{
%\begin{itemize}
%    \item Introduction: 600 words
%    \item DFT - AFM GS: 700 words
%    \item Finite T calculations: 700 words 
%    \item Topology: 700 words
%    \item Conclusion (Discussion and Outlook): 400 words
%\end{itemize}}

\section{ Introduction} 
Antiferromagnetic (AFM) metals are essential for next-generation spintronic devices due to their resilience to magnetic field disruptions, lack of stray fields, and ultrafast magnetisation dynamics~\cite{Wadley_2016, Zelezny_2014, Zelezny_2017, Roy_2016, Bodnar_2020}. Unlike conventional ferromagnetic (FM) memory devices, which are limited to gigahertz operation, the exchange-enhanced magnetisation dynamics in AFM metals can push device speeds into the terahertz regime, enabling ultrafast memory and signal processing technologies~\cite{Kamil_2018, Siddiqui_2020, Kampfrath_2011}. Additionally, AFM metals with van der Waals (vdW) stacking offer an ideal platform for electrically tunable magnetism due to their exfoliability and mobile charge carriers\cite{Shiming_2020, Bhatti_2017}. Moreover, symmetry-protected surface states in certain topological AFM materials can support dissipationless spin transport, offering novel functionalities for low-power spintronics\cite{Otrokov2019, Smejkal2022}.

\begin{figure}[t!]
    \centering
    \includegraphics[width=.9\linewidth]{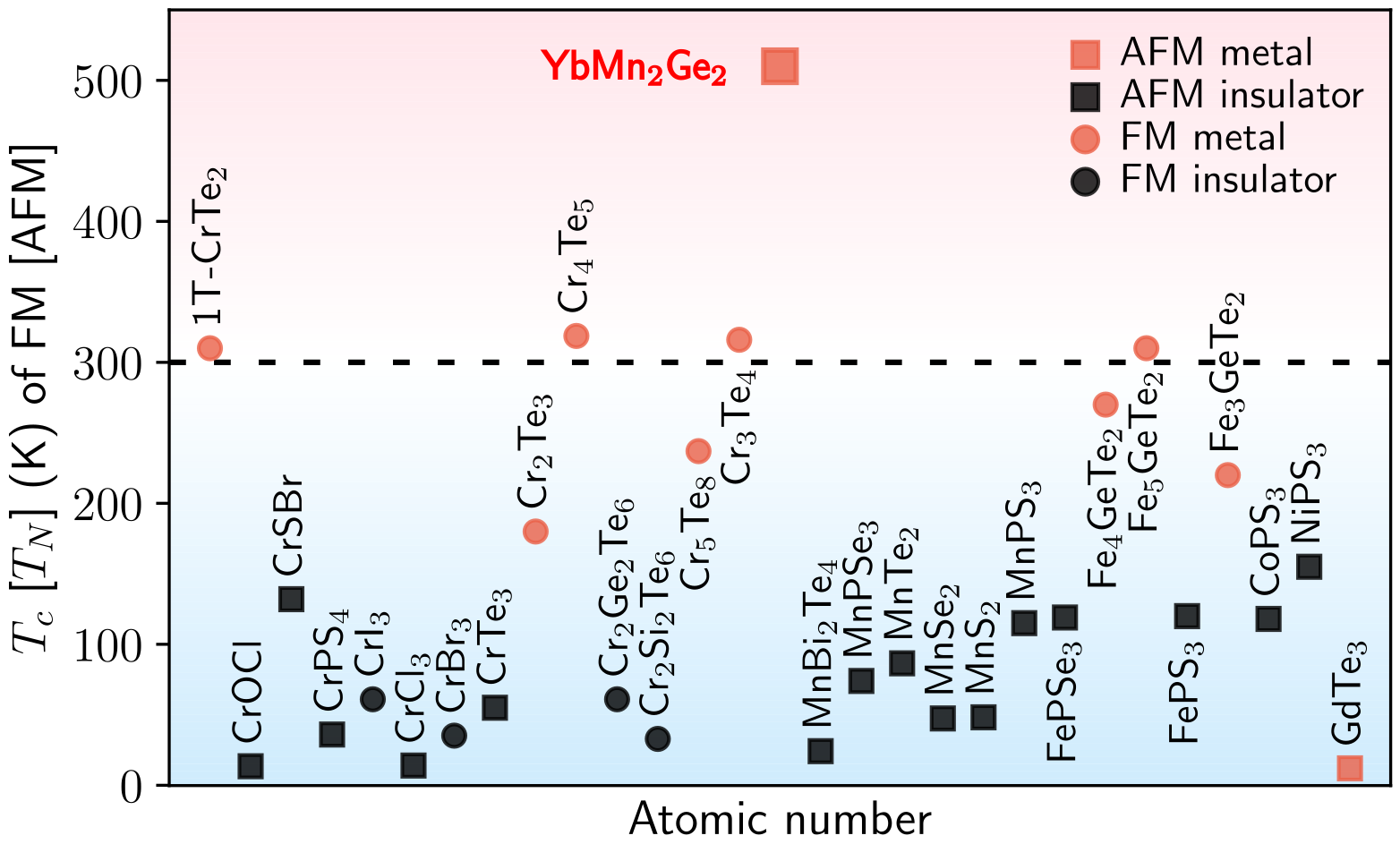}
    \caption{
    Magnetic transition temperatures of existing bulk van der Waals FM (Curie temperature) and AFM (N\'{e}el temperature) metals and insulators. Topological YMG is the only layered metallic AFM above room temperature, with a N\'{e}el temperature of $T_{N} > 500$ K. We used data from Refs.~\cite{1t-crte2_1, crocl_1, crocl_2, crsbr_1, crsbr_2, crsbr_3, crps4_1, crps4_2, cri3_1, cri3_2, crcl3_1, crbr3_1, crbr3_2, crbr3_3, crte3_1, cr2te3_1, cr2te3_2, cr2te3_3, cr4te5_1, cr2ge2te6_1, cr2ge2te6_2, cr2si2te6_1, cr2si2te6_2, cr5te8_1, cr5te8_2, cr3te4_1, cr3te4_2, Hofmann2000, mnbi2te4_1, mnpse3_1_fepse3_1, mnpse3_2_mnps3_2, mns2_1_mnse2_1_mnte2_1, mnte2_2, mnse2_2, mnse2_3, mns2_2, mns2_3, mnps3_1_feps3_1, fe4gete2_1, fe5gete2_1, fe5gete2_2, feps3_2, fe3gete2_1, fe3gete2_2, cops3_1, cops3_2, nips3_1, nips3_2, gdte3_1, gdte3_2} for the transition temperatures.
    \label{fig.1}}
\end{figure}

Despite these advantages, the realisation of room-temperature metallic and topological AFM materials remains a challenge. Strong electronic correlations in 3d transition metal compounds typically favour non-metallic ground states by penalising double electron occupancy, leading to the formation of localised moments and long-range magnetic order~\cite{Mott_1970, Cyrot_1972, Peng_2013, Takayoshi_2014, Nakatsuji_2000}. For such systems, the half-filled one-orbital Hubbard model predicts a transition from a nesting-induced antiferromagnetic Slater insulator at weak interactions to a Mott insulator as Coulomb repulsion increases~\cite{McWhan_1970, Pathak_2009, Kang_2017, Fisher_1968}. This trend, observed across numerous layered AFM materials (see Fig.~\ref{fig.1}), raises an important question: How does YbMn$_2$Ge$_2$ (YMG), a known room-temperature AFM~\cite{Nakama_2009, Qiao2021}, defy this conventional trend and retain its metallic nature?

Here, we establish the microscopic origin of metallicity in YMG and demonstrate that it is a layered exfoliable topological Dirac metal with alternagnetic surface states. Using multi-orbital Hubbard model calculations, we show that YMG’s metallic AFM state arises from a delicate balance between electronic correlations and partial Fermi surface nesting, which stabilises AFM order without fully opening a charge gap. Our first-principles calculations further reveal that YMG is easily exfoliable, making it an ideal platform for electrically tunable AFM spintronics. While the bulk of YMG is a topological metallic AFM, its surface states host a d-wave-like spin-polarised ordering. This d-wave surface spin texture resembles time-reversal symmetry broken alternagnetic features, characterised by spin-polarised bands despite vanishing net magnetisation. Our findings establish YMG as an exceptional AFM metal that simultaneously exhibits correlation effects, exfoliability, and topological properties. This provides a versatile platform for investigating the interplay of magnetism, topology, and spin transport in vdW AFMs.

\begin{figure*}
    \centering
    \includegraphics[width=1\linewidth]{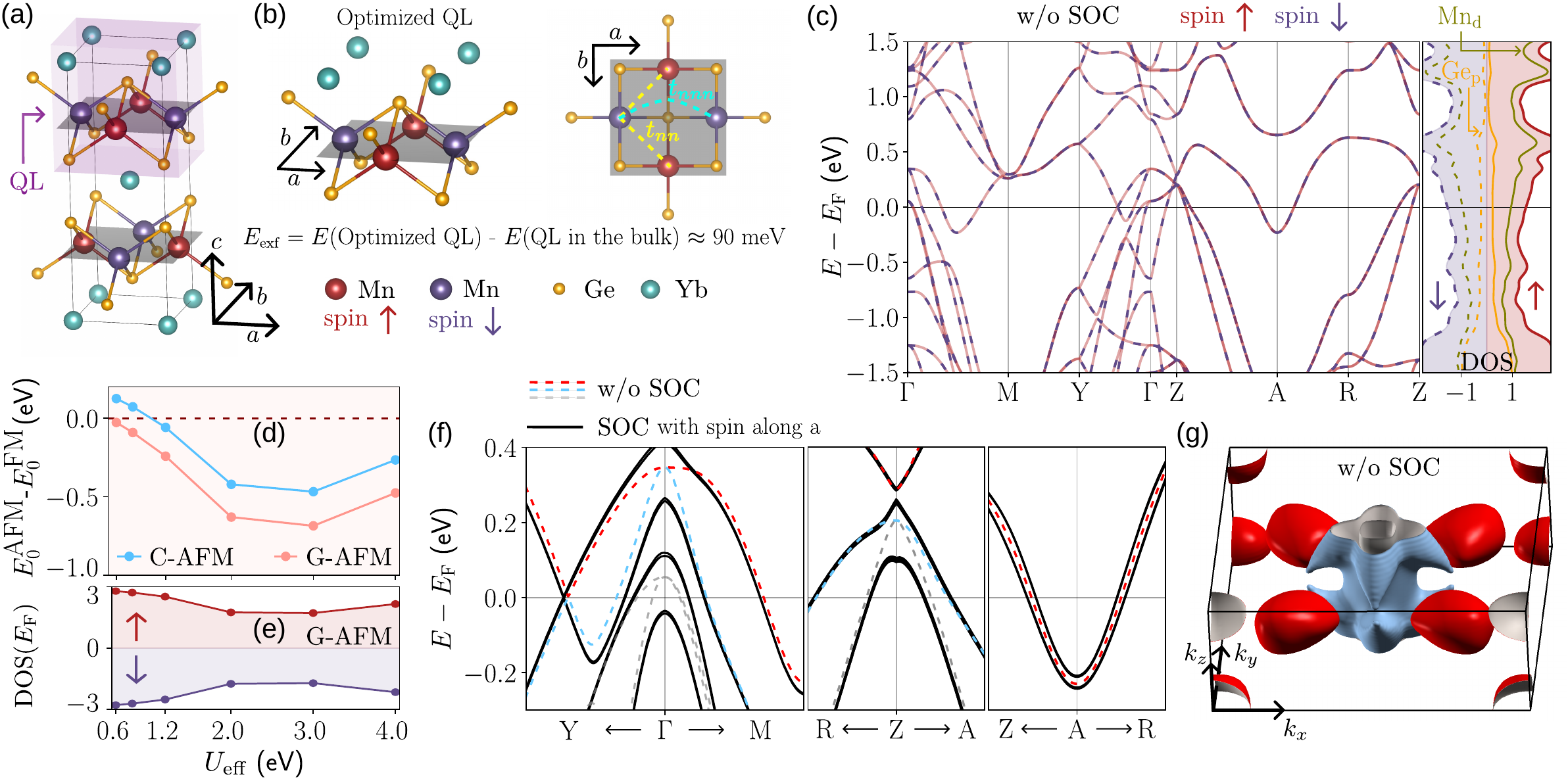}
    \caption{(a) Bulk unit cell of magnetic YMG in the G-AFM ground state. The purple box highlights the quadruple layer (QL), forming a formula unit. Consecutive QLs stacked along the crystallographic $c$-axis interact via weak vdW interaction. (b) The QL unit cell of an exfoliated 2D YMG layer in the $a-b$ plane. The relaxed 2D QL unit cell gains $\approx$ 90 meV of energy when separated from the bulk, indicating its exfoliability. The magnetic Mn atoms form a square planar network in the $ab$ plane. (c) Bulk energy dispersion along high symmetry paths and the density of states (DOS) for the G-AFM ground state of YMG in the absence of spin-orbit coupling (SOC). The presence of finite DOS at $E_{\mathrm{F}}$ establishes the metallic nature of YMG in its G-AFM ground state. (d) Variation of the energy difference of the two AFM configurations relative to the FM configuration with changing Coulomb repulsion strength $U_{\rm eff}$. The G-AFM state is the ground state for a wide range of values of Hubbard $U_{\rm eff}$ for YMG. (e) Finite DOS at the Fermi energy over a wide range of $U_{\mathrm{eff}}$ values confirms the robustness of the metallic AFM ground state. (f) The band dispersion without (dashed) and with SOC (solid black). The bands marked in red and blue dominate the DOS at $E_{\mathrm{F}}$, giving rise to the metallic nature of YMG. The SOC perturbatively affects the electronic structure by opening a small gap between the two Kramer's degenerate doublets, keeping the metallic nature intact. (g) The Fermi surfaces (FS) over the Brillouin zone.}
    \label{fig.2}
\end{figure*}

%{\it Electronic properties and AFM Ground State:---} 
\section{Metallic AFM Ground State in layered YMG}
%(Currently 750 words - 18th May) 
Bulk YMG crystallises in the tetragonal I4/mmm space group \cite{Hofmann2000}. The unit cell has a quadruple-layered (QL) structure stacked along the \textit{c} axis, with a centre of inversion at Yb, as shown in Fig.~\ref{fig.2}a. {The size of the Yb atoms was intentionally reduced in the figure to enhance the visual clarity and highlight the QL structure. 
Adjacent QL layers interact via weak vdW forces. Additionally, owing to the large interlayer super-exchange path of approximately 10.86 {\AA} between the QLs, the magnetic exchange interaction between them is also negligible. 
% Adjacent QL layers interact primarily via weak van der Waals forces. Moreover, the large interlayer superexchange path (~10.86 {\AA}) suppresses the magnetic exchange interactions between the layers significantly, making them negligible.
} 
%Additionally, this helps us highlight the strongly bonded planar network of the transition metal Mn atoms with Ge ligands in a tetrahedral environment. 
To quantify the exfoliability of YMG, we calculate its exfoliation energy ($E_{\rm exf}$) \cite{jung_2018}, the energy required to isolate a relaxed 2D YMG QL (see Fig.~\ref{fig.2}b) from the bulk structure [$E_{\rm exf}$ = $E_0$(Optimised 2D QL in free space) - $E_0$(QL in the bulk), where $E_0$ is the ground state energy]. 
For the AFM configuration of Mn atoms in a 2D QL, we find $E_{\rm exf}\approx~ 90$ meV, which is well below the exfoliation energy scale threshold of $130$ meV \cite{mounet_2018}, confirming YMG as an exfoliable layered material. To investigate the magnetic ordering in YMG, we performed \textit{ab initio} calculations (see Sec. I and II of the SI for computational details~\cite{Note1}) comparing the energy of various FM and AFM ground state configurations using GGA + U. Our calculations reveal that the lowest energy ground state is the AFM state with a G-AFM configuration, i.e. all the nearest neighbour Mn atoms are aligned anti-parallel as illustrated in Fig.~\ref{fig.2}a. 
%These results are consistent with experiments showing that YMG exhibits an AFM order~\cite{Hofmann2000, Qiao2021, Kumar2013}. 
%with a high N\'{e}el temperature $T_{\mathrm{N}}\sim$510K

In the absence of  SOC, the spin state features just the two up- and down-spin states. Therefore, we treat the non-SOC magnetic symmetry by considering the ``black-and-white group"  \cite{osti_6413225}. Our symmetry analysis shows that the two opposite magnetic sub-lattices of adjacent QL layers are related by combined parity time-reversal ($\mathcal{P}\mathcal{T}$) symmetry, %. This implies that in its G-AFM state, YMG belongs to type-III magnetic point group ~\cite{Smejkal2020}, 
with Kramer's degenerate electronic bands of the spin up and down channels.  %throughout the Brillouin zone (BZ). 
We present the band dispersion and the density of states (DOS) for the ground state magnetic configuration in Fig.~\ref{fig.2}c. Here, we set the Coulomb correlation strength for the magnetic Mn atoms to $U_{\mathrm{eff}}=3.0$~eV. 
The bands cross $E_{\mathrm{F}}$ with a large in-plane bandwidth along the $k_x-k_y$ plane ($\Gamma \rm{MY}\Gamma$), while having a much smaller bandwidth along the out-of-plane $\Gamma\rm{Z}$ direction. The in-plane bandwidth is approximately 2.4 times larger than the out-of-plane bandwidth around the Fermi energy (see SI-Fig.~2 and Sec.~II of SI for details \cite{Note1}). The anisotropy originates from strong in-plane electron hopping via the strongly bonded transition metal Mn$^{3+}$ and ligand Ge$^{4-}$ network (as shown in charge density overlap in Sec.~II of SI). In contrast, the weak charge density overlap of the ligand Ge with Yb suppresses out-of-plane hopping. Our results indicate that YMG is metallic with a finite DOS at the Fermi energy $E_{\mathrm{F}}$, primarily contributed by Mn-d orbitals, with some hybridisation from Ge-p states.

\begin{figure*}[t]
    \centering
    \includegraphics[width=0.9\linewidth]{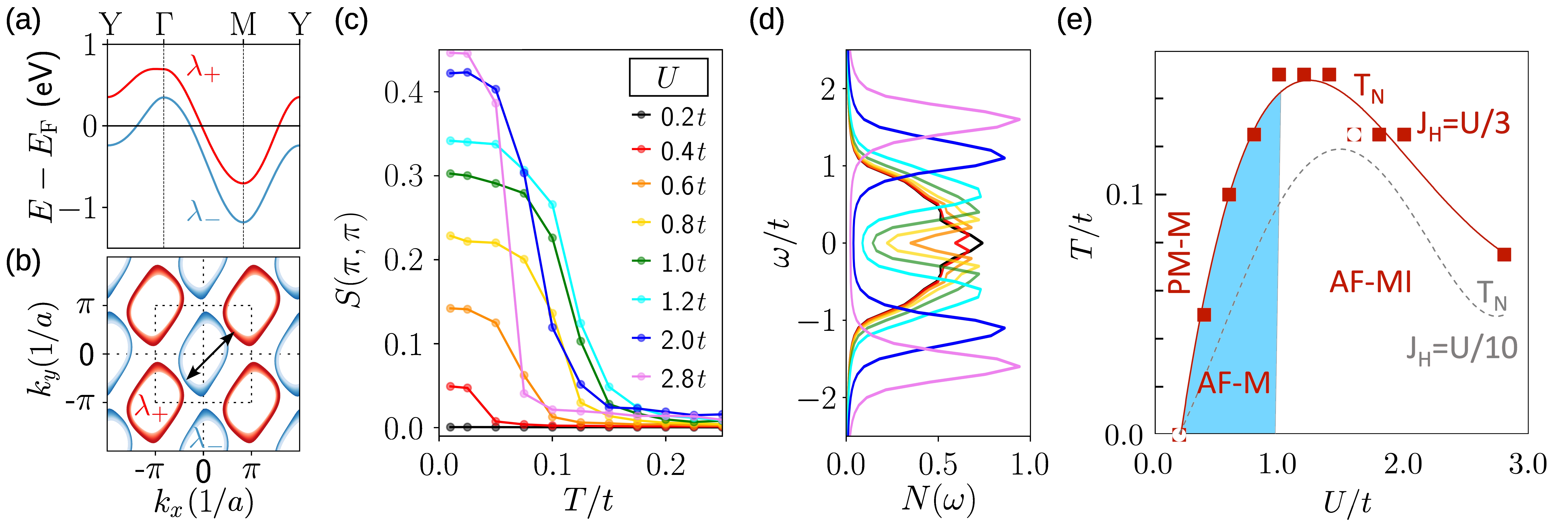}
    \caption{(a) Energy dispersion of the effective two-band TB model along high symmetry lines. (b) Contour plot of energy dispersion within an energy window of -0.12 to -0.22 eV, showing partial nesting of the red and blue bands by a vector ($\pi, \pi$). (c) Evolution of the spin structure factor with temperature for a range of $U$ values, indicating the possibility of high-temperature AFM ordering. (d) Corresponding DOS [N($\omega$)] at low temperature for the same set of $U$ values used in (c), supporting a metallic AFM state for $U/t < 1$. (e) The finite temperature phase diagram was calculated using the semi-classical Monte Carlo approach. The cyan region highlights the parameter regime supporting the AFM metal (AF-M) state, nestled between the paramagnetic metal (PM-M) and AFM Mott insulating (AF-MI) phases.}
    \label{fig.3}
\end{figure*}

 Our \textit{ab initio} calculations predict the magnetic moment to be 3.64 $\mu_B$ per Mn$^{3+}$ site, while the rest of the moment lies in the ligand sites (0.05 $\mu_B$/Ge). This is also consistent with the experimentally observed effective moment, $\mu_{\mathrm{eff}}=3.03 \mu_B$~\cite{Hofmann2000}. We confirm the robustness of the metallic AFM state of YMG with varying Coulomb correlation ($U_{\rm eff}$) by comparing the energies of different FM and AFM configurations in Fig.~\ref{fig.2}d. %We find that the G-AFM state remains the unique ground state over a wide range of $U_{\rm eff}$ values. 
Furthermore, the considerable DOS at $E_{\mathrm{F}}$ for the G-AFM state, shown in Fig.~\ref{fig.2}e for a wide range of $U_{\mathrm{eff}}$ values, 
%values of $0.6~{\rm eV} < U_{\mathrm{eff}} < 4.0$~eV, 
highlights the metallic nature of YMG. This contrasts with conventional AFM ordering, which typically leads to a Mott insulator state. In Fig.~\ref{fig.2}f, we highlight the electronic states in the vicinity of $E_{\mathrm{F}}$. The DOS at $E_{\mathrm{F}}$  has significant contributions from the lowest conduction bands (marked in red) and the highest valence bands (marked in blue), with a relatively small hole pocket from other valence bands around the $\rm Z$ point. Fig.~\ref{fig.2}g presents the Fermi surface plot in the Brillouin zone (BZ). %highlighting the contributions of the prominent bands marked in red, blue, and grey in panel (f). 

Including SOC in our GGA+SOC+U calculations reveals that the magnetic anisotropy of the AFM ordering in YMG is in-plane, consistent with experimental reports~\cite{Qiao2021}. We find that the ground state energy is identical for the N\'{e}el vector pointing along the [100], [110], and [010]  directions. The magnetic order with {in-plane [100]} spin quantization belongs to the $mm^\prime m$ magnetic point group (MPG), which includes symmetry operations $\{\mathcal{E}, C_{2y}, \mathcal{M}_{x}, \mathcal{M}_z\} + \mathcal{T}\{\mathcal{P}, C_{2x}, C_{2z}, \mathcal{M}_y\}$. Consequently, the ground state of YMG remains $P\mathcal{T}$ symmetric even in the presence of SOC. We compare the relativistic band dispersion (solid black line) for spin quantisation in the [100] direction with the non-relativistic (dashed lines) energy bands near $E_{\mathrm{F}}$ in Fig.~\ref{fig.2}f. The impact of SOC on the energy dispersion of YMG is small and perturbative, as expected in 3d transition metal-based systems~\cite{almo, Kundo2020G, Kumar2023S}. 
Having established the metallic AFM ground state in YMG, we now investigate whether this metallicity and AFM ordering persist at elevated temperatures. This is crucial for determining its potential as a robust room-temperature AFM metal. 
%Having established YMG as a G-AFM metal at low temperatures, we now focus on the temperature dependence of the metallic AFM ground state. 

%This suggests that the interplay of Coulomb interactions and multi-band physics are the primary factors dictating the metallic AFM state in YMG.

\begin{figure*}
    \centering
    \includegraphics[width=0.9\linewidth]{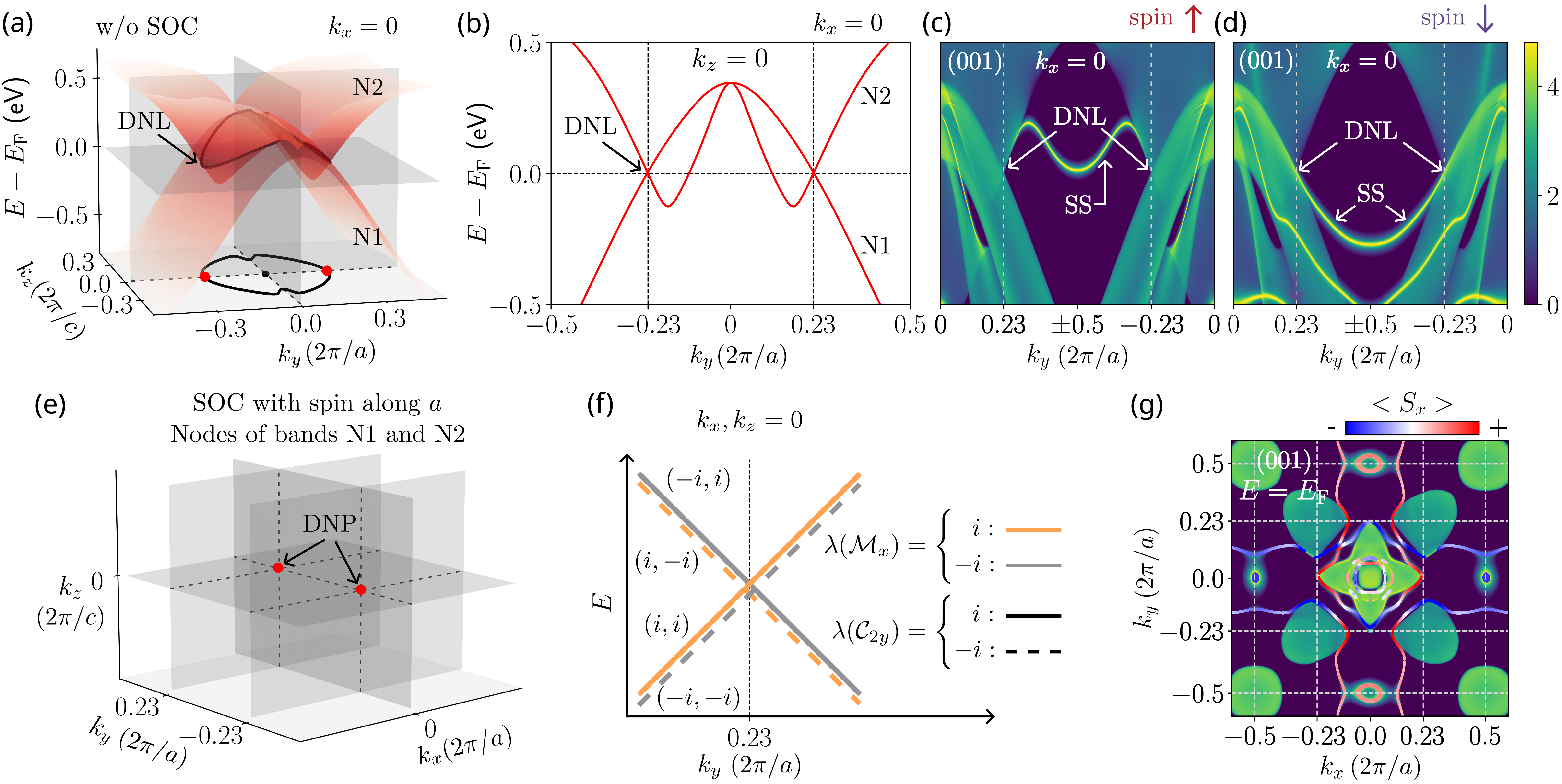}
     \caption{(a) Band dispersion of the lowest two $\mathcal{PT}$ doublets forming the nodal lines in the $k_{x}=0$ plane in the absence of SOC. The four-fold degenerate Dirac nodal lines are highlighted in black. The projection of the nodal lines on the $k_y-k_z$ plane is also shown, along with an isolated Dirac point. 
    (b) Corresponding band dispersion along $k_y$ for $k_x=k_z=0$, confirming the two nodal line Dirac crossings and an additional degenerate point at $\Gamma$. 
    (c, d) Spin-resolved spectral function of the (001) Mn terminated surface without SOC, showing both bulk and surface states. Along the $k_y$ direction, the up-spin (down-spin) channel supports a surface state with an inverted Mexican-hat like dispersion (parabolic dispersion). 
    (e) The presence of SOC (with magnetic anisotropy along the [100] direction) gaps out the nodal lines, leaving behind a pair of Dirac crossings on the $k_y$ axis, pinned at the Fermi energy. 
    (f) Each of the four bands around the symmetry-protected Dirac point has a different set of eigenvalues ($\lambda$) of ($\mathcal{M}_{x}$, $\mathcal{C}_{2y}$). 
    (g) The Fermi surface spectrum on the (001) surface, with SOC. The bulk states are in green, and we project the spin texture of the surface states with blue for down-spin and red for up-spin channel for  $\braket{S_x}$. In contrast to bulk, the surface layer breaks the $\mathcal{PT}$ symmetry and supports spin-polarised surface states. Interestingly, these surface states feature $\cal T$ broken even spin-order in momentum space.
    }
    \label{fig.4}
\end{figure*}

\section{Origin of room temperature metallic AFM state} % 880 words - As of 18th May 2024
To calculate the high-temperature phase diagram of YMG, we use a many-body Hubbard model on a square lattice system. While a full multi-orbital Hubbard model calculation for the finite temperature phase would be ideal, its computational complexity makes it infeasible. Instead, we adopt a two-orbital model that captures the dominant low-energy physics, particularly the essential features of the Fermi surface nesting that influence magnetic stability. The band-dispersion in Fig.~\ref{fig.2}f suggests that we can construct an effective two-orbital model based on the red and blue bands with dominant DOS at $E_{\mathrm{F}}$. 
The effective two-band model enables us to perform finite temperature semi-classical Monte Carlo calculations to determine the magnetic phase diagram of the model.
%While the 2-band model does not fully capture the entire band structure, it retains the essential features near the Fermi surface, particularly the $\pi-\pi$ band nesting, which is crucial for understanding the room-temperature metallic AFM state.
%In contrast, doing a finite temperature auxiliary field Monte Carlo calculation for the full band structure is not computationally prohibitive.}

We constructed a two-band effective tight-binding (TB) model Hamiltonian by analytically down-folding the Green's function around $E_{\mathrm{F}}$ (see Sec. III of the SI for details \footnote{The Supplementary Information discusses i) Details of {\it ab-initio} Calculations, ii) Correlation and the AFM ground state calculations, iii) Construction of minimal two-band tight-binding model, iv) Semi Classical Monte Carlo Method for the two-orbital Hubbard model, v) Symmetry protection of topological band crossings, and vi) Spectral functions along different terminations, and includes the following additional Refs.~[\onlinecite{kresse1996efficient, kresse1999ultrasoft,Perdew1996,paw94, paw99,Ryan_2004, szytula_2004,dudarev,giannozzi_2009,wann90,wanntools,zhi_2022,Mostofi2014, Pizzi2020,souza_2001,Noce_2014,Attias1997,Paiva_2001, Bhattacharyya_2021, Tiwari_2014,Patel_2017, Chakraborty_2022,Singh_2022}].}). The effective two-orbital nearest neighbour TB model on a square lattice network of Mn [see Fig.~\ref{fig.2}b] is 
\begin{equation}
 H_{\mathrm{TB}} = \sum_{i, \alpha, \beta, \sigma} \epsilon^{\alpha \beta} d_{i \alpha\sigma}^{\dagger}d_{i \beta\sigma} +   \sum_{\substack{\langle i,j\rangle \\ \alpha, \beta,\sigma}}\left( t_{ij}^{\alpha \beta}d_{i \alpha\sigma}^{\dagger}d_{j \beta\sigma} + h.c. \right).
 \label{eq_ham}
\end{equation}
Here, site indices $i$ and $j$ run over the nearest neighbours of the square planar Mn-network. In Eq.~\eqref{eq_ham}, $\epsilon^{\alpha \beta}$ denotes the onsite crystal field parameter and $t_{ij}^{\alpha \beta}$ is the hopping between $\alpha$ orbital of site-$i$ to $\beta$ orbital at site-$j$. % of the nearest-neighbour site connected via $a_l \hat{l}$. 
The fermionic creation (annihilation) operator,  $d^{\dagger}_{i\alpha\sigma}$ ($d_{i\alpha\sigma}$), creates (annihilates) an electron of spin $\sigma$ in orbital $\alpha$ at site $i$. We present the resultant band structure for the two bands ($\lambda_+$ in red and $\lambda_-$ in blue) along high symmetry directions of the 2D square BZ of Mn atoms in Fig.~\ref{fig.3}a. Our TB model qualitatively captures the DFT-based energy dispersion and the dominant electron and hole pockets shown in Fig.~\ref{fig.2}f. 
{The minimal model band structures along the 
$\Gamma$X direction (Fig.~\ref{fig.3}a) and the 
$\Gamma$Y direction (SI-Fig.~5b) are qualitatively similar, but show small differences due to the absence of rotational symmetry in the effective two-band model. This symmetry breaking originates from the downfolding procedure, which involves a truncated set of orbitals from the bands near the Fermi energy. We discuss this construction and its limitations in more detail in Sec.~III of the Supplementary Information\cite{Note1}.}
%
%The minimal model band structure along $\Gamma$X in Fig.~\ref{fig.3}a and $\Gamma$Y in SI-Fig.~5b are qualitatively similar, but have some differences due to the missing rotational symmetry of the original DFT band structure in the effective model. This loss of rotation symmetry in the effective two band model arises from the use of truncated set of hybridized bands in its construction. We discuss this in detail in Sec.~III of the SI \cite{Note1}.}
%In YMG bulk, hopping processes happen through the hybridised Mn d and Ge p orbitals and the charge density overlap in the two directions are symmetric. But we lose the symmetry in the process of extracting a minimal two-band model out of highly entangled bands made of many strongly hybridised orbitals, as discussed in detail in Sec. III of the SI \cite{Note1}.}

We find that the two bands ($\lambda_{+}$, $\lambda_{-}$) are nested with a wave vector ($\pi, \pi$) just below $E_{\mathrm{F}}$, as shown in Fig.~\ref{fig.3}b. Such checkerboard-shaped partial nesting is known to give rise to N\'{e}el ordering in multi-band systems \cite{Rodriguez_2018}. To calculate the finite temperature phase diagram, we use the TB model to construct a two-orbital Hubbard model for the interacting system. The details of the derivation are presented in Sec.~IV of the SI \cite{Note1}. 
%The resultant Hamiltonian is,
%\begin{eqnarray}
%H & = &   H_{\mathrm{TB}}
%+ U \sum_{i,\alpha} n_{i\alpha\uparrow}n_{i\alpha\downarrow} \nonumber \\
%&  +&~\left(U^{\prime}-\frac{J_{H}}{2}\right) \sum_{i,\alpha<\beta} n_{i\alpha}n_{i\beta} -2J_{H}\sum_{i\alpha<\beta}{S}^z_{i\alpha}{S}^z_{i\beta}  \nonumber \\
%&+ &~ J^\prime \sum_{i,\alpha<\beta}
%\left(d^\dagger_{i\alpha\uparrow}d^\dagger_{i\alpha\downarrow}
%d^{\phantom\dagger}_{i\beta\downarrow}d^{\phantom\dagger}_{i\beta\uparrow}+\text{h.c.}\right)~.
%\end{eqnarray}
The resultant Hamiltonian is $H  =   H_{\mathrm{TB}} + H_{\rm I}$, where 
\begin{eqnarray}
& & H_{\rm I}  = U \sum_{i,\alpha} n_{i\alpha\uparrow}n_{i\alpha\downarrow} 
+~\left(U^{\prime}-  \frac{J_{H}}{2}\right) \sum_{i,\alpha<\beta} n_{i\alpha}n_{i\beta} \\
& &  -  2J_{H}\sum_{i,\alpha<\beta}{S}^z_{i\alpha}{S}^z_{i\beta} 
+ J^\prime \sum_{i,\alpha<\beta}
\left(d^\dagger_{i\alpha\uparrow}d^\dagger_{i\alpha\downarrow}
d^{\phantom\dagger}_{i\beta\downarrow}d^{\phantom\dagger}_{i\beta\uparrow}+\text{h.c.}\right). \nonumber
\end{eqnarray}
The interaction Hamiltonian includes the intra-orbital ($U$) and inter-orbital ($U'$) repulsion terms along with the Hund's coupling ($J_H$) term. In terms of $U$ and $J_H$, we have $U'=U-2J_H$. Here, $n_{i\alpha}=\sum_\sigma n_{i\alpha\sigma}$ is the number operator and $S^z_{i,\alpha}=\frac{1}{2}\sum_{a,b}d^\dagger_{i\alpha,a}\hat{\sigma}^z_{a,b}d_{i\alpha,b}$.

We calculate the finite temperature phase diagram of the two-orbital Hubbard model using the recently developed semi-classical Monte Carlo (s-MC) approach. The s-MC approach allows us to access a wide temperature window and provides a reasonable phase diagram of the multi-orbital Hubbard model \cite{Anamitra_2014, Jana_2022}. For more details of the s-MC approach, see Sec.~IV of the SI \cite{Note1}. To examine the magnetic order and metallicity in YMG, we track the magnetic structure factor at various $q$ values, $S(q)$, along with the single-particle density of states, $N(\omega)$, as a function of temperature and $U$. Our structure factor analysis reveals that the magnetic order is staggered with $q=(\pi,\pi)$, consistent with the partial $(\pi,\pi)$ nesting of the Fermi surface seen in Fig.~\ref{fig.3}b. We present the evolution of the structure factor with temperature for different $U$ and $J_H=U/3$ values in Fig.~\ref{fig.3}c. The corresponding $N(\omega)$ at low temperature for the same $U$ values is shown in Fig.~\ref{fig.3}d. We find that for small $U/t<1$, the system is metallic with a finite $N(0)$ at $E_{\mathrm{F}}$, capturing the AFM metal phase, while for larger $U/t>1$, a gap develops, indicating a Mott insulating AFM phase.

We present the N\'{e}el temperature $T_N$, as a function of $U$ for two values of $J_H$ in Fig.~\ref{fig.3}e. The overall $T_N$ profile increases with increasing $J_H$.   
Figs.~\ref{fig.3}c, \ref{fig.3}d, and \ref{fig.3}e highlight that the $(\pi,\pi)$ metallic AFM state persists for a large regime of $U$ and $T$ values. Using the bandwidth of the down-folded bands in Fig.~\ref{fig.3}a, we estimate $t \sim 1.5$ eV. Combining this with the experimentally observed $T_N \sim 510$ K ~\cite{Hofmann2000, Qiao2021, Kumar2013}, we have $T/t \approx 0.03$. This corresponds to $U/t$ values of $0.3$ and $0.45$ for $J_H=U/3$ and $J_H=U/10$, respectively. Both these $U/t$ values lie in the blue region of Fig.~\ref{fig.3}e, capturing the metallic AFM state. Our s-MC results strongly support our zero temperature {\it ab initio} calculations and highlight that the metallic G-AFM state in YMG persists beyond the room temperature. The AFM metal state of YMG is reminiscent of the collinear AFM metal found in the two-orbital models of LaFeAsO, the parent compound of pnictide superconductors \cite{Raghu_2008, Anamitra_2016}. 

The Fermi surface nesting enhances the density of states near $E_{\mathrm{F}}$, increasing the exchange coupling, which helps stabilise the long-range AFM order at higher temperatures. At the same time, the partial nesting helps in sustaining metallicity by preventing a correlation induced complete gap opening at the Fermi level in the AFM. Given this robust metallic behaviour in the layered AFM, an important question arises: does the band structure of YMG also host topologically nontrivial states? To answer this, we now examine the topological features of YMG, focusing on its nodal line and Dirac-like states and the associated surface states. 

%\footnote{Note that the $T_N$ presented here corresponds to the high-temperature magnetic ordering scale reported in experiments.}.

\section{Nodal Line and Dirac Metal State of YMG} % 680 words  
%{\sout In addition to being a layered AFM metal at room temperature, YMG hosts exciting topological properties.} 
Our careful analysis reveals that without SOC, YMG hosts a pair of dispersive four-fold degenerate Dirac nodal lines between the red conduction and blue valence bands shown in Fig.~\ref{fig.2}f. We present the band-dispersion in the $k_x=0$ or, $k_y-k_z$ plane, highlighting the nodal lines and their projection on the $k_x=0$ plane (in black) in Fig.~\ref{fig.4}a. The two nodal points on the $k_x, k_z = 0$ line, which are parts of the nodal line, are pinned at $E_{\mathrm{F}}$ along the $\Gamma-\mathrm{Y}$ line, as shown in Fig.~\ref{fig.4}b. 
The identical nodal line structure for the $k_x-k_z$ plane is presented in SI-Fig.~7 of the SI \cite{Note1}. The band structure is symmetric in the $k_x-k_y$ plane, and by interchanging $k_y$ and $k_x$, we obtain plots identical to Fig.~\ref{fig.4}a and b for the $k_x$ direction (see SI-Fig.~7).

To identify the origin of the nodal lines, we consider the minimal set of four bands needed for a Dirac crossing. The generic low-energy Hamiltonian describing such four bands is given by \cite{Tang2016, fu_kane_2007},
\begin{equation}
    H(\textbf{k}) = d_{0}(\textbf{k})~\mathcal{I}_{4\times 4} + \sum_{i=1,2,3,4,5}d_{i}(\textbf{k})\Gamma_{i}~.
\end{equation}
Here, the five Dirac matrices, $\Gamma_{1} = \tau_{x}\otimes \sigma_{0}$, $\Gamma_{2} = \tau_{z}\otimes \sigma_{0}$, $\Gamma_{3} = \tau_{y}\otimes \sigma_{x}$, $\Gamma_{4} = \tau_{y}\otimes \sigma_{y}$, $\Gamma_{5} = \tau_{y}\otimes \sigma_{z}$ are constructed from the Pauli matrices $\tau_\alpha $ and $\sigma_\alpha$ with $\alpha = \{x, y, z\}$. The Pauli matrices represent the orbital ($\tau_\alpha $) and the spin ($\sigma_\alpha$) degrees of freedom. The coefficients $d_i(\textbf{k})$ (with $i = \{0,\cdots, 5\}$) are real functions. All four bands become degenerate at a $\textbf{k}$ point if $d_{i}(\textbf{k}) = 0$ for all $i$. Below, we analyse the impact of crystalline symmetry operations of magnetic YMG on the nodal line and band crossing points captured by this model. 

The G-AFM ground state of YMG hosts two mirror symmetries, $\mathcal{M}_{x}$ and $\mathcal{M}_{y}$. 
%\st{The $\mathcal{M}_x$ is defined in the invariant $k_x = 0$ plane, based on its eigenvalue equation and the commutation relation with $\mathcal{PT}$.}
On the {$k_x = 0$ plane,} the coefficients $d_i(0, k_y, k_z)$ remain invariant under the operation of $\mathcal{M}_{x}$. At the same time, the Dirac matrices $\Gamma_{1}$ and $\Gamma_{3}$ anti-commute with  $\mathcal{M}_{x}$ (see Sec.~V of SI for details \cite{Note1}). Combining this with the invariance of the Hamiltonian on the mirror symmetry-protected $k_{x} = 0$ plane forces the coefficients $d_{1}$ and $d_{3}$ to vanish (see Sec.~V of the SI for explicit derivation \cite{Note1}). Consequently, in the $k_x = 0$ plane, 
dispersive nodal lines are allowed if $d_{2}(k_{x}=0, k_{y}, k_{z}) = 0$, $d_{4}(k_{x}=0, k_{y}, k_{z}) = 0$, and $d_{5}(k_{x}=0, k_{y}, k_{z}) = 0$. This establishes that the nodal lines in YMG are accidental and not enforced by crystalline symmetry. Additionally, we find that the two bands forming the Dirac nodal line have opposite parity eigenvalues under $\mathcal{M}_x$, allowing the bands to cross each other, as shown in SI-Fig. 6 of the SI \cite{Note1}. Similarly, nodal line solutions can emerge on the $k_{y}=0$ plane due to the presence of $\mathcal{M}_{y}$ symmetry. We present the up- and down-spin channel spectral functions of bulk and surface states for the easily cleavable (001) surface in Figs.~\ref{fig.4}c and \ref{fig.4}d, respectively. Isolated and dispersive surface states for the up-spin channel (down-spin channel) with an inverted Mexican-hat like dispersion (parabolic dispersion) can be clearly seen along the $k_y$ direction. The corresponding plot for the spectral function in the $k_x$ direction is presented in SI-Fig.~7 of Sec.~V of SI \cite{Note1}.  

In presence of SOC with magnetisation anisotropy along the $a$ axis \cite{Qiao2021}, the $\mathcal{M}_{x}$ symmetry remains intact, while the $\mathcal{M}_{y}$ symmetry is broken. The anisotropy of the spin orientation in the orbital space leads to different crystalline symmetries, compared to the no SOC scenario (see Sec.~V of SI for details~\cite{Note1}).  Furthermore, YMG contains an additional two-fold $\mathcal{C}_{2y}$ rotation symmetry with the spin anisotropy along $a$. The $\mathcal{C}_{2y}$ symmetry forces the $d_{4}(k_{x} = 0, k_{y}, k_{z} = 0)$ coefficient of the Hamiltonian to be zero on the $k_y$ axis (see Sec.~V of the SI for details~\cite{Note1}). As discussed above, the $d_{1}$, and $d_{3}$ coefficients vanish on the $k_x = 0$ plane owing to the ${\cal M}_x$ symmetry. Combining these restrictions, Dirac crossing can appear only when $d_{2}(k_{x} = 0, k_{y}, k_{z} = 0)$, and $d_{5}(k_{x} = 0, k_{y}, k_{z} = 0)$ is zero on the $k_y$ axis. Accordingly, we find two stable isolated Dirac crossings at $k_y=\pm 0.23$, equidistant from the $\Gamma$ point on the $k_y$ axis [see Fig.~\ref{fig.4}e]. Both Dirac points are pinned at $E_{\mathrm{F}}$. 
%We highlight that the SOC enforced in-plane magnetocrystalline anisotropy alters the MPG symmetries of YMG, leading to the Dirac nodal loop to Dirac node pair transition. 
We highlight that the inclusion of SOC alters the symmetries of YMG, which leads to a Dirac nodal loop to Dirac crossings transition {\cite{Tang2016, Thakur_2017, Bahadur_2018, bahadur_2018_starfruit}}. 

In Fig.~\ref{fig.4}f, we show that the four bands forming the Dirac crossing carry different sets of eigenvalues for the $\mathcal{M}_{x}$ and $\mathcal{C}_{2y}$ symmetry operators. This stabilises the Dirac crossings and protects them from level repulsion. Dirac Fermions are known to coexist with AFM ordering in  AFe$_2$As$_2$ (A=Ba, Sr) \cite{basrfe2as2_topological_Dirac} and in CuMnAs \cite{smejkal_2017_prl}, as well. In YMG, the two bands are gapped at all other $k$-points in the $k$-space as shown in the nodal plot in Fig.~\ref{fig.4}e. The $\mathcal{Z}_2$ number on the $k_y = 0$ and $k_y = \pi/a$ planes are 0 and 1 respectively. That captures the non-trivial topological phase transition along the $k_y$ direction at the Dirac crossing in YMG, which is shown in the evolution of the wannier charge centre in Sec. V of the SI \cite{Note1}. The bulk topology leads to non-trivial surface states. We present the spin-projected spectral function of the Mn-terminated (001) surface along with the bulk states over $k_x-k_y$ plane in Fig.~\ref{fig.4}g. Owing to ${\cal P}{\cal T}$ symmetric Kramer's degenerate bulk bands, the bulk spectral functions feature no spin-splitting. However, the (001) surface breaks the $\mathcal{PT}$ symmetry, and as a consequence, the surface states become spin-polarised. Remarkably, the spin polarisation of  YMG surface states is distinct from the spin-polarised surface states observed in other Dirac systems such as CuMnAs \cite{Tang2016}, 
% NiTe$_2$ \cite{Ghosh2019O}, PtTe$_2$\cite{Yan_2017, ptte2_barun}, and CoTe$_2$\cite{cote2_advanced_materials, cote2_atasi} 
MTe$_2$ (M=Ni\cite{Ghosh2019O}, Pt\cite{Yan_2017, ptte2_barun}, Co\cite{cote2_advanced_materials, cote2_atasi}). More interestingly, the in-plane nearest neighbour surface Mn atoms are connected by $C_{4z}\mathcal{T}$ symmetry. Consequently, the spin-polarised surface features an even d-wave spin-polarised order in the momentum space (see Fig.~\ref{fig.4}g). 
%To the best of our knowledge, this is the first report of such d-wave altermagnetic spin-polarized surface states in any material. \textcolor{red}{--AC: There are 2D d-wave altermagnet which has similar symmetries. so we can tone-down this.} 

%Note that SOC did not influence the high-temperature metallic AFM state or the characteristics of the surface states of YMG. However, the soc-enforced in-plane magnetocrystalline anisotropy alters the MPG symmetries of YMG, leading to the Dirac nodal loop to Dirac node pair transition.}
%significant changes in the topology and surface states of YMG.

% \textcolor{red}{
% % Similar AFM topological Dirac cones are seen in AFM AFe$_2$As$_2$ (A=Ba, Sr) \cite{basrfe2as2_topological_Dirac}. Symmetry-protected Dirac crossing and gap opening due to symmetry breaking are also shown in CuMnAs \cite{smejkal_2017_prl}. 
% The presence of Fermi arc states is also observed in AFM NdBi\cite{schrunk_2022}. The symmetry protection of surface Dirac crossing by $t\mathcal{T}$, where $t$ is translation and $\mathcal{T}$ is time-reversal symmetries, as shown for NdBi \cite{honma_2023}, is different from our system.}
% % We don't have $t\mathcal{T}$ in our system and the symmetry protection is governed by different crystal symmetries in our $\mathcal{PT}$ protected system.}

\section{Conclusion}% 180 words 
Despite their vast advantages in AFM spintronics and memory device applications, room-temperature metallic AFM candidates are scarce. We have demonstrated that YMG is an exfoliable layered antiferromagnetic Dirac metal at room temperature. Combining our {\it ab-initio} calculations with the semi-classical Monte Carlo approach, we show that the AFM metal state in YMG arises from the interplay of short-range correlation effects and the partial ($\pi, \pi$) nesting of the Fermi surface. Furthermore, the symmetry-protected topological Dirac crossings in bulk give rise to novel spin-polarised surface states with altermagnetic d-wave like spin texture, with no net magnetisation. 

Our findings motivate further investigations into transport phenomena, spintronic applications, and emergent physics at the intersection of magnetism, topology, and metallicity in YMG and related compounds. For example, the preserved ${\cal P} {\cal T}$ symmetry in bulk YMG suggests the potential realization of both dissipationless and dissipative intrinsic nonlinear Hall and longitudinal transport phenomena\cite{Chakraborty_2022,debottam_2024, kamal_shibalik_2023}, which could be leveraged for energy-efficient device applications. Further exploration of YMG and related van-der-Waals materials~\cite{Hofmann2001, basrfe2as2_topological_Dirac, MALAMAN1994209} could unlock new pathways for designing power-efficient, fast-switching spintronic devices. 

\section*{Acknowledgement}
We acknowledge the high-performance computing facility at IIT Kanpur for computational support. NJ thanks Debasis Dutta for technical discussions. AC acknowledges financial support from Alexander von Humboldt postdoctoral fellowship. AM acknowledges the use of the NOETHER high-performance cluster at NISER. 

\bibliography{reference}

\end{document}